\documentstyle[eqsecnum,pre,aps,epsf,epsfig]{revtex}

\begin{document}
\draft

\title{Nonlinear random resistor diode networks and fractal dimensions of directed percolation clusters
}
\author{Olaf Stenull and Hans-Karl Janssen
}
\address{
Institut f\"{u}r Theoretische Physik 
III\\Heinrich-Heine-Universit\"{a}t\\Universit\"{a}tsstra{\ss}e 1\\40225 
D\"{u}sseldorf, Germany
}
\date{\today}
\maketitle

\begin{abstract}
We study nonlinear random resistor diode networks at the transition from the non percolating to the directed percolating phase. The resistor-like bonds and the diode-like bonds under forward bias voltage obey a generalized Ohm's law, $V \sim I^r$. Based on general grounds as symmetries and relevance we develop a field theoretic model. We focus on the average two-port resistance, which is governed at the transition by the resistance exponent $\phi_r$. By employing renormalization group methods we calculate $\phi_r$ for arbitrary $r$ to one-loop order. Then we address the fractal dimensions characterizing directed percolation clusters. Via considering distinct values of the nonlinearity $r$, we determine the dimension of the red bonds, the chemical path and the backbone to two-loop order.
\end{abstract}
\pacs{PACS numbers: 64.60.Ak, 05.60.-k, 72.80.Ng}

%\twocolumn
\section{Introduction}
Percolation\cite{bunde_havlin_91_etc} describes the passage of an influence through a medium which is irregularly structured in the sense that the influence can propagate through some regions whereas it cannot pass other areas. Prominent examples for such media are computer networks like the internet where information propagates and irregularity can be caused by random switch failures or other technical problems.

A particularly simple percolation model is the random resistor network (RRN). In this model the irregular medium is given by a, say hypercubic, lattice in which bonds between nearest neighboring sites are randomly occupied with a probability $p$. The influence corresponds to an external current $I$, which is injected at a terminal site $x$ and withdrawn at another terminal site $x^\prime$. Depending on the occupation probability $p$ the resistors (bonds) are likely to either be isolated or to form clusters. Two sites belong to the same cluster if they are connected by a path of bonds and hence current can flow between them. At low  $p$ two infinitely separated terminal sites $x$ and $x^\prime$ are not connected by such a path and the network behaves as an insulator. For large $p$, on the other hand, many paths between $x$ and $x^\prime$ may exist and the network is a conductor. Therefore, at some probability in between, a threshold $p_c$ must exist where for the first time current can percolate from $x$ to $x^\prime$. The threshold probability is called the percolation threshold. Since it separates the conducting and the insulating phase, it is also referred to as the critical probability. In RRNs the influence can percolate through occupied bonds in all directions. The resulting clusters are typically isotropic in space. This kind of percolation is referred to as isotropic percolation (IP). The linear extend of the isotropic clusters can be characterized by the correlation length $\xi \sim \left| p - p_c \right|^{-\nu}$, where $\nu$ is the correlation length exponent of the IP universality class.

Directed percolation (DP)\cite{hinrichsen_2000} is an anisotropic variant of percolation. The bonds function as diodes so that the current can percolate only along a given distinguished direction. The critical properties of isotropic and directed percolation are very different. Typical DP clusters are anisotropic and they are characterized by two different correlation lengths: $\xi_{\parallel}$ (parallel to the distinguished direction) and $\xi_\perp$ (perpendicular to it). As one approaches the critical probability, the two correlation lengths diverge with the exponents $\nu_\parallel$ and $\nu_\perp$ of the DP universality class.

The apparent success of DP might be attributed to the fact that it is perhaps the simplest model resulting in branching self-affine objects. It has many potential applications, including fluid flow through porous media under gravity, hopping conductivity in a strong electric field\cite{vanLien_shklovskii_81}, crack propagation\cite{kertez_vicsek_80}, and the propagation of surfaces at depinning transitions\cite{depinning}.

DP has a dynamic interpretation in which the distinguished direction is viewed as  time. A DP cluster then represents the history of a stochastic process. In this dynamic interpretation the DP universality class is the generic universality class for phase transitions from an active to an absorbing inactive state. For example the epidemic spreading of an infectious desease without immunization\cite{murray_88} may be described by DP\cite{grassberger_85}. Moreover, DP is related to self-organized critical models~\cite{soc}.

In the early 1980's Redner\cite{red_81&82a,red_83,perc} introduced the random resistor diode network (RDN) which comprises both, IP and DP. A RDN is a bond percolation model where nearest-neighbor sites are connected by a resistor, a positive diode (conducting only in the distinguished direction), a negative diode (conducting only opposite to the distinguished direction), or an insulator with respective probabilities $p$, $p_{+}$, $p_{-}$, and $q=1-p-p_{+}-p_{-}$. In the three dimensional phase diagram (pictured as a tetrahedron spanned by the four probabilities) one finds a nonpercolating and three percolating phases. The percolating phases are isotropic, positively directed, or negatively directed. Between the phases there are surfaces of continuous transitions. All four phases meet along a multicritical line, where $0\leq r:=p_{+}=p_{-}\leq 1/2$ and $p=p_{c}(r)$. On the entire multicritical line, i.e., independently of $r$, one finds the scaling properties of usual isotropic percolation ($r=0$). For the crossover from IP to DP see, e.g., Ref.\cite{janssen_stenull_2000}. In this paper we focus exclusively on the vicinity of the critical surface separating the non-percolating and the positively directed phase.

An important notion in the theory of RDN is the average resistance $M_R (x, x^\prime )$ between two connected terminal sites $x = (x_\perp , t)$ and $x^\prime = (x_\perp^\prime , t^\prime)$ of the network. The critical behavior of this average resistance is well known\cite{janssen_stenull_directedLetter_2000,stenull_janssen_directedResistance_2000}. If $M_R (x, x^\prime )$ is measured for example in the distinguished, timelike direction then it scales like
\begin{eqnarray}
M_R \left( x, x^\prime \right) \sim \left| t - t^\prime \right|^{\phi /\nu_\parallel} \ ,
\end{eqnarray}
with $\phi$ being the DP resistance exponent. Here in this paper we consider a generalized RDN in which  the resistor-like bonds and the diode-like bonds under forward bias voltage obey a generalized Ohm's law, $V \sim I^r$. Our motivation to assume this non linear Ohm's law is twofold. First, real circuit elements have non linear current-voltage characteristics. This is obviously the case for diodes but it also applies to a certain extend to resistors, in particular for large currents. Our main motivation is, however, that the generalized average resistance $M_{R_r} (x, x^\prime )$ is related for certain values of $r$ to important fractal substructures of DP clusters. This relation provides us with an elegant way to determine the fractal dimensions of the red bonds, the chemical path, and the DP backbone. Parts of this work have been presented briefly in Ref.~\cite{janssen_stenull_directedLetter_2000}.

The plan of presentation is the following. In Sec.~\ref{theModel} we give background on non linear RDN and set up our field theoretic model. Section~\ref{rga} sketches our renormalization group improved perturbation calculation. We derive the scaling behavior of $M_{R_r} (x, x^\prime )$ which is governed by a generalized resistance exponent $\phi_r$. We present our one-loop result for $\phi_r$. In Sec.~\ref{fracDim} we calculate the fractal dimensions of the red bonds, the chemical path, and the DP backbone by considering the limits $r \to \infty$, $r \to 0^+$, and $r \to -1^+$, respectively. Section~\ref{concls} contains our conclusions.

\section{The model}
\label{theModel}

\subsection{Generalized Ohm's law}
Consider a $d$-dimensional hypercubic lattice in which the direction ${\rm{\bf n}} = 1/\sqrt{d} \left( 1, \dots , 1 \right)$ is distinguished. Assume that the bonds $\underline{b}_{\langle i,j \rangle}$ between two nearest neighboring sites $i$ and $j$ are directed so that $\underline{b}_{\langle i,j \rangle} \cdot {\rm{\bf n}} > 0$.

Suppose that the directed bonds obey the non-linear Ohm's law\cite{kenkel_straley_82}
\begin{mathletters}
\label{ohmR}
\begin{eqnarray}
\sigma_{\underline{b}_{\langle i,j \rangle}} \left( V_{\underline{b}_{\langle i,j \rangle}} \right) \, V_{\underline{b}_{\langle i,j \rangle}} \, \left| V_{\underline{b}_{\langle i,j \rangle}} \right|^{s-1} = I_{\underline{b}_{\langle i,j \rangle}} \ ,
\end{eqnarray}
or, equivalently,
\begin{eqnarray}
V_{\underline{b}_{\langle i,j \rangle}} = \rho_{\underline{b}_{\langle i,j \rangle}} \left( I_{\underline{b}_{\langle i,j \rangle}} \right) \, I_{\underline{b}_{\langle i,j \rangle}} \, |I_{\underline{b}_{\langle i,j \rangle}}|^{r-1} \ .
\end{eqnarray}
\end{mathletters}
$V_{\underline{b}_{\langle i,j \rangle}} = V_j - V_i$, where $V_i$ denotes the potential at site $i$, is the voltage drop over the bond between sites $i$ and $j$. $I_{\underline{b}_{\langle i,j \rangle}}$ denotes the current flowing from $j$ to $i$. In the following we drop the subscript $\langle i,j \rangle$ whenever it is save. The bond conductances $\sigma_{\underline{b}}$ are random variables taking on the values $\sigma$, $\sigma \theta \left( V \right)$, $\sigma \theta \left( -V \right)$, and $0$ with respective probabilities $p$, $p_+$, $p_-$, and $q$. $\sigma$ is a positive constant and $\theta$ denotes the Heaviside function. The exponents $r$ and $s$ are describing the non linearity with $r=s^{-1}$. The bond resistances $\rho_{\underline{b}}$ are related to the conductances via $\sigma_{\underline{b}} = \rho_{\underline{b}}^{-s}$. Note that the diodes are idealized: under forward-bias voltage they behave for $r=1$ as ``ohmic'' resistors whereas they are insulating under backward-bias voltage. We point out that the round brackets in Eq.~(\ref{ohmR}) contain the argument of the bond conductance or resistance, respectively. It is important to realize that these quantities depend on the voltages or currents by means of the step function and that $\mbox{sign} \left(  V_{\underline{b}} \right) = \mbox{sign} \left(  I_{\underline{b}} \right)$. Hence we may write $\sigma_{\underline{b}} \left( V_{\underline{b}} \right) = \sigma_{\underline{b}} \left( I_{\underline{b}} \right)$ and $\rho_{\underline{b}} \left( I_{\underline{b}} \right) = \rho_{\underline{b}} \left( V_{\underline{b}} \right)$.

Assume that an external current $I$ is injected at $x$ and withdrawn at $x^\prime$. It is understood that $x$ and $x^\prime$ are connected. The power dissipated on the network is by definition 
\begin{eqnarray}
P=I \left( V_x - V_{x^\prime} \right) \ .
\end{eqnarray}
Using Ohm's law it may be expressed entirely in terms of the voltages as
\begin{eqnarray}
\label{power1}
P = R_{r,+} (x ,x^\prime)^{-1} \left| V_x - V_{x^\prime} \right|^{s+1} = \sum_{\underline{b}} \sigma_{\underline{b}} \left( V_{\underline{b}} \right) \left| V_{\underline{b}} \right|^{s+1} = P \left( \left\{ V \right\} \right) \ .
\end{eqnarray}
The sum is taken over all bonds on the cluster connecting $x$ and $x^\prime$ and $\left\{ V \right\}$ denotes the corresponding set of voltages. $R_{r,+} (x ,x^\prime)$ stands for the macroscopic resistance when $I$ is inserted at $x$ and withdrawn at $x^\prime$. Similarly one may define $R_{r,-} (x ,x^\prime)$ as the macroscopic resistance when $I$ is inserted at $x^\prime$ and withdrawn at $x$. The two quantities are related by $R_{r,+} (x ,x^\prime) = R_{r,-} (x^\prime ,x)$. From the power one obtains Kirchhoff's first law
\begin{eqnarray}
\label{kirchhoff}
\sum_{\langle j \rangle} \sigma_{\underline{b}_{\langle i,j \rangle}} \left( V_{\underline{b}_{\langle i,j \rangle}} \right) V_{\underline{b}_{\langle i,j \rangle}} \left| V_{\underline{b}_{\langle i,j \rangle}} \right|^{s-1} = \sum_{\langle j \rangle} I_{\underline{b}_{\langle i,j \rangle}} = - I_i 
\end{eqnarray}
as a consequence of the variation principle
\begin{eqnarray}
\label{variationPrinciple1}
\frac{\partial}{\partial V_i} \left[ \frac{1}{s+1} P \left( \left\{ V \right\} \right) + \sum_k I_k V_k \right] = 0 \ .
\end{eqnarray}
The summation in Eq.~(\ref{kirchhoff}) extends over the nearest neighbors of $i$ and $I_i$ is given by $I_i = I \left( \delta_{i,x} - \delta_{i,x^\prime} \right)$.

Alternatively to Eq.~(\ref{power1}), the power can be expressed in terms of the currents as
\begin{eqnarray}
\label{power2}
P = R_{r,+} \left( x ,x^\prime \right) \left| I \right|^{r+1} = \sum_{\underline{b}} \rho_{\underline{b}} \left( I_{\underline{b}} \right) \left| I_{\underline{b}} \right|^{r+1} = P \left( \left\{ I \right\} \right) \ ,
\end{eqnarray}
with $\left\{ I \right\}$ denoting the set of currents flowing through the individual bonds. It is understood that $\rho_{\underline{b}} \left( I_{\underline{b}} \right) \left| I_{\underline{b}} \right|^{r+1} = 0$ whenever $\sigma_{\underline{b}} \left( I_{\underline{b}} \right) = 0$. Kirchhoff's second law, saying that the voltage drops along closed loops vanish, can be stated in terms of the variation principle\begin{eqnarray}
\label{variationPrinciple2}
\frac{\partial}{\partial I^{(l)}} P \left( \left\{ I^{(l)} \right\} , I \right) = 0 \ ,
\end{eqnarray}  
i.e., there are no independent loop currents $I^{(l)}$ circulating around a complete set of independent closed loops.

\subsection{Relation to fractal substructures}
\label{clusterProperties}
Now we take a short detour and show that the average resistance $M_{R_r} (x, x^\prime )$ is related for specific values of $r$ to the mass, i.e., the average number of bonds, of important substructures of DP clusters. The arguments below are well established for RRN and we simply adapt them here to RDN. We start with the backbone. The (forward) backbone between two sites $x$ and $x^\prime$ is defined, in principle, as the union of all current carrying bonds when $I$ is injected at $x$ and withdrawn at $x^\prime$. Consider $r \to -1^+$. One obtains immediately as a consequence of Eq.~(\ref{power2}), that
\begin{eqnarray}
\label{rTo-1}
R_{-1,+} (x ,x^\prime) = \lim_{r\to -1^+} \sum_{\underline{b}} \rho_{\underline{b}} \left( I_{\underline{b}} \right) \left| 
\frac{I_{\underline{b}}}{I}\right|^{r+1} = \sum_{\underline{b}} \rho_{\underline{b}} \left( I_{\underline{b}} \right)\ ,
\end{eqnarray}
with only those bonds carrying non zero current contributing to the sum on the right hand side. Hence 
\begin{eqnarray}
\label{massBB}
\lim_{r\to -1^+} M_{R_r} (x ,x^\prime) \sim M_B \ ,
\end{eqnarray}
where $M_B$ stands for the mass of the backbone. 

Now we turn to $r \to \infty$ and $r \to 0^+$ following the lines of Blumenfeld and Aharony\cite{blumenfeld_aharony_85}. On the backbone between two sites $x$ and $x^\prime$ one may distinguish between two different substructures: blobs formed by multi-connected bonds and singly connected bonds which are referred to as red bonds. Both substructures are contributing to the resistance of the backbone
\begin{eqnarray}
R_{r,+} (x ,x^\prime) = \sum_{\underline{b}}^{\mbox{\scriptsize blob \normalsize}} 
\rho_{\underline{b}} \left( I_{\underline{b}} \right) \left| \frac{I_{\underline{b}}}{I}\right|^{r+1} + \sum_{\underline{b}}^{\mbox{\scriptsize red \normalsize}} \rho_{\underline{b}} \left( I_{\underline{b}} \right) \ ,
\end{eqnarray}
where the sums are taken over all bonds belonging to blobs and over all red bonds respectively. 
Since sites on a blob are multi-connected by definition, $|I_{\underline{b}}| < |I|$, and thus
\begin{eqnarray}
\lim_{r\to \infty} \sum_{\underline{b}}^{\mbox{\scriptsize blob \normalsize}} 
\rho_{\underline{b}} \left( I_{\underline{b}} \right) \left| \frac{I_{\underline{b}}}{I}\right|^{r+1} = 0 
 \ .
\end{eqnarray}
In conclusion, $M_{R_{r}} (x ,x^\prime)$ is related to the mass of the red bonds $M_{\mbox{\scriptsize red \normalsize}}$ via
\begin{eqnarray}
\label{massRed}
\lim_{r\to \infty} M_{R_r} (x ,x^\prime) \sim M_{\mbox{\scriptsize red \normalsize}} \ .
\end{eqnarray}

Consider now the first site $x$ at some end of a blob. An entering current $I$ 
splits into currents $I_{i,x}$ flowing to nearest neighbors $i$ with
\begin{eqnarray}
\left| I_{i,x} \right| = \sigma_{i,x} \left( V_x - V_i \right) \, \left| V_x - V_i \right|^s  \ .
\end{eqnarray}
In the limit $s \to \infty$ the ratios $|I_{i,x}| / |I_{j,x}|$ vanish whenever $\sigma_{i,x} \left( V_x - V_i \right) \, \left| V_x - V_i \right|^s < \sigma_{j,x} \left( V_x - V_j \right) \, \left| V_x - V_j \right|^s$. Thus, current flows only through the resistor with the largest $\sigma_{i,x} \left( V_x - V_i \right) \,\left| V_x - V_i \right|^s$. This argument may be iterated through the entire blob. One identifies either a single self avoiding chain through which $I$ flows, with  
\begin{eqnarray}
\label{powerOfChain}
P_r = \sum_{\underline{b}} \rho_{\underline{b}} \left( I_{\underline{b}} \right) |I|^{r+1} 
\end{eqnarray}
being the power dissipated on the chain, or several of such chains with identical power. The expression in Eq.~(\ref{powerOfChain}) is minimal for minimal $\sum_{\underline{b}} \rho_{\underline{b}}$, i.e., the current chooses the shortest path through the blob and one is led to
\begin{eqnarray}
\label{dMinRel}
\lim_{r\to 0^+} M_{R_{r}} (x ,x^\prime) \sim M_{\mbox{{\scriptsize min}}} \ , 
\end{eqnarray}
where $M_{\mbox{{\scriptsize min}}}$ stands for the mass of the chemical path.

\subsection{Replica technique}
\label{replicaFormalism}
Our aim now is to calculate the average resistance between two ports $x$ and $x^\prime$ which is precisely defined by
 \begin{eqnarray}
M_{R_r} (x ,x^\prime) = \langle \chi_+ (x ,x^\prime) R_{r,+} (x ,x^\prime ) \rangle_C / \langle \chi_+ (x ,x^\prime) \rangle_C \ . 
\end{eqnarray}
$\langle ...\rangle_C$ denotes the average over all configurations of the 
diluted lattice. $\chi_+ (x ,x^\prime)$ is an indicator function that takes the value one if $x$ and $x^\prime$ are positively connected, i.e., if $I$ can percolate from $x$ to $x^\prime$, and zero otherwise. Note that $\langle \chi_+ (x ,x^\prime) \rangle_C = \langle \chi_- (x^\prime ,x) \rangle_C$ is nothing more than the usual DP correlation function.

Now we follow an idea by Stephen\cite{stephen_78} and its generalization to networks of nonlinear resistors by Harris\cite{harris_87} and exploit correlation functions 
\begin{eqnarray}
\label{defCorr}
G \left( x, x^\prime ,\vec{\lambda} \right) = \left\langle 
\psi_{\vec{\lambda}}(x)\psi_{-\vec{\lambda}}(x^\prime) 
\right\rangle_{\mbox{\scriptsize{rep}}}
\end{eqnarray}
of
\begin{eqnarray}
\label{defPsi}
\psi_{\vec{\lambda}}(x) = \exp \left( i \vec{\lambda} \cdot \vec{V}_x \right) \ , \quad \vec{\lambda} \neq \vec{0}\ ,
\end{eqnarray}
as generating functions of $M_{R_r}$. In writing Eqs.~(\ref{defCorr}) and (\ref{defPsi}) we switched to $D$-fold replicated voltages, $V_i \to \vec{V_i} = \left( V_i^{(1)}, \cdots , V_i^{(D)} \right)$, and imaginary currents, $\lambda_i = i I_i \to \vec{\lambda_i} = \left( \lambda_i^{(1)}, \cdots , \lambda_i^{(D)} \right)$. The correlation functions 
are given by
\begin{eqnarray}
\label{erzeugendeFunktion}
G \left( x, x^\prime ,\vec{\lambda} \right) &=& \Bigg\langle Z^{-D} \int \prod_j \prod_{\alpha =1}^D dV_j^{(\alpha )} \exp \Bigg[ -\frac{1}{s+1} P \left( \left\{ \vec{V} \right\} \right) 
\nonumber \\
&& + \, \frac{i \omega}{2} \sum_i \vec{V}_i^2 + i \vec{\lambda} \cdot \left( 
\vec{V}_x  - \vec{V}_{x^\prime} \right) \Bigg] \Bigg\rangle_C \ ,
\end{eqnarray}
where
\begin{eqnarray}
\label{repPower}
P \left( \left\{ \vec{V} \right\} \right) = \sum_{\alpha =1}^D P \left( \left\{ V^{(\alpha )} \right\} \right) = \sum_{\alpha =1}^D \sum_{\underline{b}} \sigma_{\underline{b}} \left(  V_{\underline{b}}^{(\alpha )} \right) \left| V_{\underline{b}}^{(\alpha )} \right|^{s+1}
\end{eqnarray}
and $Z$ is the normalization
\begin{eqnarray}
\label{norm}
Z = \int \prod_{j} dV_{j} \exp \left[ -\frac{1}{s+1} P \left( \left\{ V \right\} \right) + \frac{i \omega}{2} \sum_i V_i^2 \right] \ .
\end{eqnarray}
The additional power term $\frac{i\omega}{2} \sum_i V^2_i$ which we have introduced  in Eqs.~(\ref{erzeugendeFunktion}) and (\ref{norm}) is necessary to give the voltage integrals a well defined meaning. Without this term the integrands depend only on voltage differences and the integrals are divergent. Physically the new term corresponds to grounding each lattice site by a capacitor of unit capacity. The original situation can be retrieved by taking the limit of vanishing frequency, $\omega \to 0$.

Because the integrations in Eqs.~(\ref{erzeugendeFunktion}) and (\ref{norm}) are not Gaussian, we employ the saddle point method. The saddle point equation is nothing more than the variation principle stated in Eq.~(\ref{variationPrinciple1}). Thus, the maximum of the integrand is determined by the solution of the circuit equations (\ref{kirchhoff}). Provided that the condition $|I|^{r+1} \gg \sigma$ holds, we obtain, up to an unimportant multiplicative constant which goes to one in the limit $D \to 0$,
\begin{eqnarray}
\label{GenFkt}
G \left( x, x^\prime ,\vec{\lambda} \right) = \left\langle \exp \left[  
\frac{\Lambda_r \left( \vec{\lambda} \right)}{r+1} R_{r,+} \left( x,x^\prime \right)  \right] \right\rangle_C \ ,
\end{eqnarray}
where
\begin{eqnarray}
\Lambda_r \left( \vec{\lambda} \right)=\sum_{\alpha =1}^D \left( - i \lambda^{(\alpha )} 
\right)^{r+1} \ .
\end{eqnarray}
Now we may expand $G$ about $\Lambda_r \left( \vec{\lambda} \right) =0$, 
\begin{eqnarray}
\label{expansionOfG}
G \left( x, x^\prime ,\vec{\lambda} \right) = \left\langle \chi_+ ( x,  x^\prime ) \right\rangle_C \left( 1 + \frac{\Lambda_r \left( \vec{\lambda} \right)}{r+1} M_{R_r} ( x, x^\prime ) + \cdots 
\right) \ .
\end{eqnarray}
This shows us that $G$ is indeed the desired generating function from which the average resistance may be calculated via
\begin{eqnarray}
\label{shit}
M_{R_r} ( x, x^\prime ) = \left\langle \chi_+ ( x,  x^\prime ) \right\rangle_C^{-1} \frac{\partial}{\partial \left( \Lambda_r \left( \vec{\lambda} \right) /(r+1) \right)} G \left( x, x^\prime ,\vec{\lambda} \right) \Big|_{\Lambda_r \left( \vec{\lambda} \right) =0} \ .
\end{eqnarray}

Here we would like to comment on the nature of $\vec{\lambda}$. We work near the limit when all the components of $\vec{\lambda}$ are equal and continue to large imaginary values. Accordingly we set\cite{harris_87}
\begin{eqnarray}
\label{crazyLambda}
\lambda^{(\alpha )} = i \lambda_0 + \xi^{(\alpha )} \ ,
\end{eqnarray}
with real $\lambda_0$ and $\xi^{(\alpha )}$, and impose the condition $\sum_{\alpha =1}^D \xi^{(\alpha )} = 0$. The saddle point approximation may be justified by demanding
\begin{eqnarray}
\label{cond1}
\left| \lambda_0 \right| \gg 1 . 
\end{eqnarray}

Substitution of Eq.~(\ref{crazyLambda}) into the definition of $\Lambda_r$ leads to
\begin{eqnarray}
 \Lambda_r \left( \vec{\lambda} \right) &=& \sum_{\alpha =1}^D \left\{ \lambda_0^{r+1} - i \left( r+1 \right) \lambda_0^{r} \xi^{(\alpha )} - \frac{r \left( r+1 \right) }{2} \lambda_0^{r-1} 
\xi^{(\alpha ) 2} + \cdots \right\}
\nonumber \\
&=&
D  \lambda_0^{r+1} - \frac{r \left( r+1 \right) }{2} \lambda_0^{r-1} \vec{\xi}^2 
+ \cdots \ .
\end{eqnarray}
Thus, one can justify the expansion in Eq.~(\ref{expansionOfG}) by invoking the 
conditions
\begin{eqnarray}
\label{cond2}
\left| \lambda_0 \right|^{r+1} \ll D^{-1} \quad \mbox{and} \quad \left| \lambda_0 \right|^{r-1} \vec{\xi}^2 
\ll 1 \ .
\end{eqnarray}
Note that the replica limit $D\to 0$ allows for a simultaneous fulfillment of the conditions (\ref{cond1}) and (\ref{cond2}). However, we will not only rely exclusively on these conditions on $\vec{\lambda}$. We will provide several consistency checks for the validity of Harris' saddle point approach as we go along and reproduce known results.

\subsection{Field theoretic Hamiltonian}
Since infinite voltage drops between different clusters may occur, it is not guaranteed that $Z$ stays finite, i.e., the limit $\lim_{D \to 0}{Z^D}$ is not well defined. This problem can be regularized by switching to voltage variables $\vec{\vartheta}$ taking discrete values  on a $D$-dimensional torus~\cite{harris_lubensky_87}. The voltages are discretized by setting $\vec{\vartheta} = \Delta \vartheta \vec{k}$, where $\Delta \vartheta = \vartheta_M /M$ is the gap between successive voltages, $\vartheta_M$ is a voltage cutoff, $\vec{k}$ is a $D$-dimensional integer, and $M$ a positive integer. The components of $\vec{k}$ are restricted to $-M < k^{(\alpha)} \leq M$ and periodic boundary conditions are realized by equating $k^{(\alpha )}=k^{(\alpha )} \mbox{mod} (2M)$. The continuum may be restored by taking $\vartheta_M \to \infty$ and $\Delta \vartheta \to 0$. By setting $\vartheta_M = \vartheta_0 M$, $M=m^2$, and, respectively, $\Delta \vartheta = \vartheta_0 /m$, the two limits can be taken simultaneously via $m \to \infty$. Note that the limit $D \to 0$ has to be taken before any other limit. Since the voltages and $\vec{\lambda}$ are conjugated variables, $\vec{\lambda}$ is affected by the discretization as well:
\begin{eqnarray}
\vec{\lambda} = \Delta \lambda \, \vec{l} \ , \ \Delta \lambda \, \Delta \vartheta = \pi /M \ ,
\end{eqnarray}
where $\vec{l}$ is a $D$-dimensional integer taking the same values as $\vec{k}$. This choice guarantees that the completeness and orthogonality relations
\begin{mathletters}
\label{complete}
\begin{eqnarray}
\frac{1}{(2M)^D} \sum_{\vec{\vartheta}} \exp \left( i \vec{\lambda} \cdot \vec{\vartheta} \right) = \delta_{\vec{\lambda} ,\vec{0} 
\hspace{0.15em}\mbox{\scriptsize{mod}}(2M \Delta \lambda) }
\end{eqnarray}
and
\begin{eqnarray}
\frac{1}{(2M)^D} \sum_{\vec{\lambda}} \exp \left( i \vec{\lambda} \cdot \vec{\vartheta} \right) = \delta_{\vec{\vartheta} ,\vec{0} 
\hspace{0.15em}\mbox{\scriptsize{mod}}(2M \Delta \vartheta)}
\end{eqnarray}
\end{mathletters}
do hold. Equation~(\ref{complete}) provides us with a Fourier transform between the $\vec{\vartheta}$- and $\vec{\lambda}$-tori.

After taking care of these regularization issues, we now carry out the average over the diluted lattice configurations in Eq.~(\ref{erzeugendeFunktion}). This provides us with the effective Hamiltonian
\begin{eqnarray}
\label{effHamil}
H_{\mbox{\scriptsize{rep}}} &=&  - \ln \left\langle  \exp \left[ - \frac{1}{s+1} P \left( \left\{ \vec{\vartheta} \right\} \right) + \frac{i\omega}{2} \sum_i \vec{\vartheta}_{i}^2 \right] \right\rangle_C 
\nonumber \\
&=& - \sum_{\underline{b}} K \left( \vec{\vartheta}_{\underline{b}} \right) - \frac{i\omega}{2} \sum_i \vec{\vartheta}_{i}^2 \ ,
\end{eqnarray}
where
\begin{eqnarray}
\label{kern1}
K \left( \vec{\vartheta}\right) &=& \ln \bigg\{ q + p \prod_{\alpha =1}^D \exp \left[ - \frac{\sigma}{s+1} \left| \vartheta^{(\alpha )} \right|^{s+1} \right]
+ p_+ \prod_{\alpha =1}^D \exp \left[ - \frac{\sigma}{s+1} \theta \left( \vartheta^{(\alpha )} \right) \left| \vartheta^{(\alpha )} \right|^{s+1} \right] 
\nonumber \\
&+& p_- \prod_{\alpha =1}^D \exp \left[ - \frac{\sigma}{s+1} \theta \left( - \vartheta^{(\alpha )} \right) \left| \vartheta^{(\alpha )} \right|^{s+1} \right] \bigg\} \ .
\end{eqnarray}
In order to proceed further we recall the choice for $\vec{\lambda}$ made in Eq.~(\ref{crazyLambda}). Because $\vec{\lambda}$ and $\vec{\vartheta}$ are related by Ohm's law, we have to make a consistent choice for $\vec{\vartheta}$:
\begin{eqnarray}
\vartheta^{(\alpha )} = \vartheta_0 + \zeta^{(\alpha )} \ ,
\end{eqnarray}
with real $\vartheta_0$ and $\zeta^{(\alpha )}$, and where $\sum_{\alpha =1}^D \zeta^{(\alpha )} = 0$. Upon imposing the condition
\begin{eqnarray}
\left| \vartheta_0 \right| \gg \left| \zeta^{(\alpha )} \right|
\end{eqnarray}
for all $\alpha$ we may write
\begin{eqnarray}
\label{kern2}
K \left( \vec{\vartheta}\right) &=& \ln \bigg\{ q + p \exp \left( - \frac{\sigma}{s+1} \left| \vartheta \right|^{s+1} \right)
+ p_+ \left[ \theta \left( -  \vartheta_0 \right) + \theta \left( \vartheta_0 \right) \exp \left( - \frac{\sigma}{s+1} \left| \vartheta \right|^{s+1} \right) \right]
\nonumber \\
&+& p_- \left[ \theta \left( \vartheta_0 \right) + \theta \left( - \vartheta_0 \right) \exp \left( - \frac{\sigma}{s+1} \left| \vartheta \right|^{s+1} \right) \right] \bigg\} \ ,
\end{eqnarray}
where we have introduced the abbreviation $\left| \vartheta \right|^{s+1} = \sum_{\alpha = 1}^D \left| \vartheta^{(\alpha )} \right|^{s+1}$. After doing a little straightforward algebra and by dropping a term
\begin{eqnarray}
\theta \left( \vartheta_0 \right) \ln \left[ 1 - p - p_+ \right] + \theta \left( - \vartheta_0 \right) \ln \left[ 1 - p - p_- \right] 
\end{eqnarray}
which does not depend on the bond conductances we obtain
\begin{eqnarray}
\label{kern3}
K \left( \vec{\vartheta}\right) = \theta \left( \vartheta_0 \right) K_+ \left( \vec{\vartheta}\right) + \theta \left( - \vartheta_0 \right) K_- \left( \vec{\vartheta}\right) \ .
\end{eqnarray}
The $K_\pm \left( \vec{\vartheta} \right)$ in Eq.~(\ref{kern3}) are given by
\begin{eqnarray}
\label{defKpm}
K_\pm \left( \vec{\vartheta}\right) = \ln \left[ 1 + \frac{p + p_\pm}{ 1 - p - p_\pm} \exp \left( - \frac{\sigma}{s+1} \left| \vartheta \right|^{s+1} \right) \right] \ .
\end{eqnarray}
Note that these are exponentially decreasing functions in replica space with a decay rate proportional to $\sigma^{-1}$. In order to refine $H_{\mbox{\scriptsize{rep}}}$ towards a field theoretic Hamiltonian we now expand $K \left( \vec{\vartheta}\right)$ in terms of $\psi$:
\begin{eqnarray}
\label{kern4}
K \left( \vec{\vartheta}_{\underline{b}} \right) &=& \frac{1}{\left( 2M \right)^D} \sum_{\vec{\lambda}} \sum_{\vec{\vartheta}} \exp \left[ i \vec{\lambda} \cdot \left( \vec{\vartheta}_{\underline{b}} - \vec{\vartheta} \right) \right] K \left( \vec{\vartheta} \right) \ .
\end{eqnarray}
Upon exploiting that $\theta \left( \vartheta_0 \right) = \theta \left( \lambda_0 \right)$ we obtain
\begin{eqnarray}
\label{kern5}
K \left( \vec{\vartheta}_{\underline{b}} \right) 
&=& \sum_{\vec{\lambda} \neq \vec{0}} \psi_{\vec{\lambda}} \left( i \right) \psi_{-\vec{\lambda}} \left( j \right) 
\bigg\{ \frac{1}{2} \left[ \widetilde{K}_+ \left( \vec{\lambda} \right) + \widetilde{K}_- \left( \vec{\lambda} \right) \right] 
\nonumber \\
&& + \frac{1}{2} \left[ \theta \left( \lambda_0 \right) - \theta \left( -\lambda_0 \right) \right] \left[ \widetilde{K}_+ \left( \vec{\lambda} \right) - \widetilde{K}_- \left( \vec{\lambda} \right) \right] \bigg\} \ .
\end{eqnarray}
where $\widetilde{K}_\pm \left( \vec{\lambda} \right)$ stands for the Fourier transform of $K_\pm \left( \vec{\vartheta}\right)$, 
\begin{eqnarray}
\widetilde{K}_\pm \left( \vec{\lambda} \right) = \frac{1}{\left( 2M \right)^D} \sum_{\vec{\vartheta}} \exp \left[ i \vec{\lambda} \cdot \vec{\vartheta} \right] K \left( \vec{\vartheta} \right) \ .
\end{eqnarray}
The Fourier transform can be carried out by switching back to continuous currents and expanding the logarithm in Eq.~(\ref{defKpm}). The so obtained result has the Taylor expansion
\begin{eqnarray}
\label{kernelExpansion}
\widetilde{K}_\pm \left( \vec{\lambda} \right) = \tau_\pm - w_{r,\pm} \, \Lambda_r \left( \vec{\lambda} \right) + \cdots \ ,
\end{eqnarray}
where $\tau_\pm$ and $w_{r, \pm} \sim \sigma^{-1}$ are expansion coefficients depending on $p$ and $p_\pm$ with $\tau_\pm (p, p_+, p_-) = \tau_\mp (p, p_-, p_+)$ and $w_{r, \pm} (p, p_+, p_-) = w_{r, \mp} (p, p_-, p_+)$. Now we insert Eq.~(\ref{kern5}) into Eq.~(\ref{effHamil}). We also carry out a gradient expansion in position space. This is justified because only nearest neighbor pairs enter in the power $P$, i.e., the interaction is short ranged not only in replica but also in position space. We find 
\begin{eqnarray}
\label{effHamil2}
H_{\mbox{\scriptsize{rep}}} &=&  - \sum_{\vec{\lambda} \neq \vec{0}} \sum_{i , \underline{b}_i} \bigg\{ \frac{1}{2} \left[ \widetilde{K}_+ \left( \vec{\lambda} \right) + \widetilde{K}_- \left( \vec{\lambda} \right) \right] \psi_{-\vec{\lambda}} \left( i \right) \left[ 1  + \frac{1}{2} \left( \underline{b}_i \cdot \nabla \right)^2 + \cdots \right] \psi_{\vec{\lambda}} \left( i \right)
\nonumber \\
&&+  \frac{1}{2} \left[ \theta \left( \lambda_0 \right) - \theta \left( -\lambda_0 \right) \right] \left[ \widetilde{K}_+ \left( \vec{\lambda} \right) - \widetilde{K}_- \left( \vec{\lambda} \right) \right] \psi_{-\vec{\lambda}} \left( i \right) \left[ \underline{b}_i \cdot \nabla + \cdots \right] \psi_{\vec{\lambda}} \left( i \right) \ ,
\end{eqnarray}
with $\widetilde{K}_\pm \left( \vec{\lambda} \right)$ given by Eq.~(\ref{kernelExpansion}).

We proceed with the usual coarse graining step and replace the 
$\psi_{\vec{\lambda}} \left( i \right)$ by order parameter fields $\psi_{\vec{\lambda}} \left( {\rm{\bf x}} \right)$ which inherit the constraint $\vec{\lambda} \neq \vec{0}$. We model the corresponding field theoretic Hamiltonian $\mathcal H \mathnormal$ in the spirit of Landau as a mesoscopic free energy and introduce the Landau-Ginzburg-Wilson type functional
\begin{eqnarray}
\label{hamiltonian}
{\mathcal{H}} &=& \int d^dx \bigg\{ \frac{1}{2} \sum_{\vec{\lambda} \neq \vec{0}} \psi_{-\vec{\lambda}} \left( {\rm{\bf x}} \right) \left[  \tau - \nabla^2 - w_r \Lambda_r \left( \vec{\lambda} \right) + \left( \theta \left( \lambda_0 \right) - \theta \left( -\lambda_0 \right) \right) {\rm{\bf v}} \cdot \nabla \right] \psi_{\vec{\lambda}} \left( {\rm{\bf x}} \right)
\nonumber \\
& & + \, \frac{g}{6} \sum_{\vec{\lambda}, \vec{\lambda}^\prime  , \vec{\lambda} + \vec{\lambda}^\prime \neq \vec{0}} \psi_{-\vec{\lambda}} \left( {\rm{\bf x}} \right) \psi_{-\vec{\lambda}^\prime} \left( {\rm{\bf x}} \right) \psi_{\vec{\lambda} + \vec{\lambda}^\prime} \left( {\rm{\bf x}} \right) + \frac{i \omega}{2} \nabla^2_{\vec{\lambda}} \psi_{\vec{\lambda}} \left( {\rm{\bf x}} \right) \bigg\} \ .
\end{eqnarray}
As usual we have neglected all terms which are irrelevant in the sense of the renormalization group. The parameter $\tau$ is the coarse grained ancestor of $\tau_+ + \tau_-$. It specifies the ``distance'' from the critical surface under consideration. $w_r \sim \sigma^{-1}$ is the coarse grained analog of $w_{r, +} + w_{r, -}$. The vector ${\rm{\bf v}}$ lies in the distinguished direction, ${\rm{\bf v}} = v {\rm{\bf n}}$. $v$ depends as $\tau$ and $w_r$ on the three probabilities $p$, $p_+$, and $p_-$. For $p_+ = p_-$ it vanishes. In the limit $w_r\to 0$ our Hamiltonian ${\mathcal{H}}$ describes the usual purely geometric DP. Indeed ${\mathcal{H}}$ leads for $w_r\to 0$ to exactly the same perturbation series as obtained in\cite{cardy_sugar_80,janssen_81,janssen_2000}. In the limit $r \to 1$ ${\mathcal{H}}$ describes resistor diode percolation as studied in Refs.~\cite{janssen_stenull_directedLetter_2000,stenull_janssen_directedResistance_2000}. In the remainder of this paper we drop the regularization term proportional to $\omega$ for simplicity.

\section{Renormalization group analysis}
\label{rga}
Now we are in the position to set up a perturbation calculation. This perturbation calculation can be simplified from the onset by manipulating ${\mathcal{H}}$ in such a way that it takes the form of a dynamic functional\cite{janssen_dynamic,deDominicis&co,janssen_92}. We assume that $v \neq 0$ and introduce new variables by setting
\begin{eqnarray}
\label{subst}
x_\parallel = {\rm{\bf n}} \cdot {\rm{\bf x}} = v \rho t \ , \quad 
\psi = |v|^{-1/2} \, s \ , \quad
g = |v|^{1/2} \, \overline{g} \ .
\end{eqnarray}
By substituting Eq.~(\ref{subst}) into Eq.~(\ref{hamiltonian}) we obtain 
\begin{eqnarray}
\label{dynFktnal}
{\mathcal{J}} &=& \int d^{d_\perp}x_\perp  \, dt \bigg\{ \frac{1}{2} \sum_{\vec{\lambda} \neq \vec{0}} s_{-\vec{\lambda}} \left( {\rm{\bf x}}_\perp , t \right) \left[ \rho \left( \tau - \nabla^2_\perp - w_r \Lambda_r \left( \vec{\lambda} \right) \right) + \left( \theta \left( \lambda_0 \right) - \theta \left( -\lambda_0 \right) \right) \frac{\partial}{\partial t} \right] s_{\vec{\lambda}} \left( {\rm{\bf x}}_\perp , t  \right)
\nonumber \\
& & + \, \frac{\rho \overline{g}}{6} \sum_{\vec{\lambda}, \vec{\lambda}^\prime  , \vec{\lambda} + \vec{\lambda}^\prime \neq \vec{0}} s_{-\vec{\lambda}} \left( {\rm{\bf x}}_\perp , t  \right) s_{-\vec{\lambda}^\prime} \left( {\rm{\bf x}}_\perp , t  \right) s_{\vec{\lambda} + \vec{\lambda}^\prime} \left( {\rm{\bf x}}_\perp , t  \right) \bigg\} \ ,
\end{eqnarray}
where $d_\perp = d-1$. Note that we have neglected a term containing a second derivative with respect to the "time" $t$. This is justified because this term is less relevant than the one with the first "time" derivative which we kept.

We proceed with standard methods of field theory\cite{amit_zinn-justin}. From Eq.~(\ref{dynFktnal}) we gather the diagrammatic elements contributing to our perturbation series. The first element is the vertex $- \overline{g}$. Dimensional analysis shows that the vertex $\overline{g}$ is marginal in four transverse dimensions. Hence $d = d_\perp +1 = 5$ is the upper critical dimension as it is well known for DP. The second diagrammatic element is the Gaussian propagator $G \left( {\rm{\bf x}}_\perp , t , \vec{\lambda} \right)$ which is determined by the equation of motion
\begin{eqnarray}
\left\{ \rho \left[ \tau - \nabla^2 - w_r \Lambda_r \left( \vec{\lambda} \right) \right] + \left[ \theta \left( \lambda_0 \right) - \theta \left( -\lambda_0 \right) \right] \frac{\partial}{\partial t} \right\} G \left( {\rm{\bf x}}_\perp , t , \vec{\lambda} \right) = \delta \left( {\rm{\bf x}}_\perp \right) \delta \left( t \right) .
\end{eqnarray}
For the Fourier transformed $\widetilde{G} \left( {\rm{\bf p}} , t , \vec{\lambda} \right)$ of $G \left( {\rm{\bf x}}_\perp , t , \vec{\lambda} \right)$, where ${\rm{\bf p}}$ is the momentum conjugate to ${\rm{\bf x}}_\perp$, one obtains readily
\begin{eqnarray}
\widetilde{G} \left( {\rm{\bf p}} , t , \vec{\lambda} \right) = \widetilde{G}_+ \left( {\rm{\bf p}} , t , \vec{\lambda} \right) + \widetilde{G}_- \left( {\rm{\bf p}} , t , \vec{\lambda} \right) \ .
\end{eqnarray}
The quantities on the right hand side are given by
\begin{eqnarray}
\label{defGpm}
\widetilde{G}_\pm \left( {\rm{\bf p}} , t , \vec{\lambda} \right) = \theta \left( \pm t \right) \theta \left( \pm \lambda_0 \right) \exp \left[ \mp t \rho \left( \tau + {\rm{\bf p}}^2 - w_r \Lambda_r \left( \vec{\lambda} \right) \right) \right] \left( 1 - \delta_{\vec{\lambda}, \vec{0}} \right) \ .
\end{eqnarray}
For the diagrammatic expansion it is sufficient to keep either $\widetilde{G}_+ \left( {\rm{\bf p}} , t , \vec{\lambda} \right)$ or $\widetilde{G}_- \left( {\rm{\bf p}} , t , \vec{\lambda} \right)$. We choose to keep $\widetilde{G}_+ \left( {\rm{\bf p}} , t , \vec{\lambda} \right)$. 

\subsection{Nonlinear resistance of Feynman diagrams}
From the vertex $- \overline{g}$ and the propagator $\widetilde{G}_+ \left( {\rm{\bf p}} , t , \vec{\lambda} \right)$ we now assemble the Feynman graphs constituting our diagrammatic expansion. As in our previous work on transport in IP\cite{stenull_janssen_oerding_99,janssen_stenull_oerding_99,janssen_stenull_99,stenull_2000,stenull_janssen_2000a} these Feynman diagrams have a real-world interpretation: they may be viewed as being directed resistor networks themselves. This real-world interpretation has basically two roots. The first one is that the principal propagator $\widetilde{G}_+ \left( {\rm{\bf p}} , t , \vec{\lambda} \right)$ decomposes into two parts:
\begin{eqnarray}
\label{decoGp}
\widetilde{G}_+ \left( {\rm{\bf p}} , t , \vec{\lambda} \right) &=& \theta \left( t \right) \theta \left( \lambda_0 \right) \exp \left[ - t \rho \left( \tau + {\rm{\bf p}}^2 - w_r \Lambda_r \left( \vec{\lambda} \right) \right) \right] 
\nonumber \\
&& - \theta \left( t \right)  \exp \left[ - t \rho \left( \tau + {\rm{\bf p}}^2 \right) \right] \delta_{\vec{\lambda}, \vec{0}} \ .
\end{eqnarray}
One of these parts is carrying $\vec{\lambda}$'s and hence we call it conducting. The other one is not carrying $\vec{\lambda}$'s and accordingly we call it insulating. Equation~(\ref{decoGp}) allows for a schematic decomposition of the principal diagrams into sums of conducting diagrams consisting of conducting and insulating propagators. In Fig.~1 we list the conducting diagrams resulting from the decomposition procedure up to two-loop order. The second root of the real-world interpretation is that the replica currents $\vec{\lambda}$ are conserved in each vertex just as currents are conserved in nodes of real networks. Hence we may write for each edge $i$ of a diagram, $\vec{\lambda}_i = \vec{\lambda}_i \left( \vec{\lambda} , \left\{ \vec{\kappa} \right\} \right)$, where $\vec{\lambda}$ is an external current and $\left\{ \vec{\kappa} \right\}$ denotes a complete set of independent loop currents. The $\vec{\lambda}$-dependent part of each conducting diagram then takes the form
\begin{eqnarray}
\label{lambdaAnteil}
\exp \left[ \rho \, w_r \sum_{i} t_i \Lambda_r \left( \vec{\lambda}_i \right) \right] \ .
\end{eqnarray} 
Now it is important to realize that $\sum_{i} t_i \Lambda_r \left( \vec{\lambda}_i \right)$ resembles the structure of a power (cf.~Eq.~(\ref{power2})). Thus, we interpret the "time" associated with a conducting propagator as its resistance and write 
\begin{eqnarray}
\mbox{(\ref{lambdaAnteil})} = \exp \left[ \rho \, w_r P_r \left( \vec{\lambda} , \left\{ \vec{\kappa} \right\} \right) \right] \ .
\end{eqnarray} 

The real-world interpretation provides for an alternative way of computing the conducting Feynman diagrams. To evaluate the sums over independent loop currents,
\begin{eqnarray}
\label{toEvaluate}
\sum_{\left\{ \vec{\kappa} \right\}} \exp \left[ \rho \, w_r P_r \left( \vec{\lambda} , \left\{ \vec{\kappa} \right\} \right) \right] \ ,
\end{eqnarray}
we employ the saddle point method under the conditions discussed at the end of Sec.~\ref{replicaFormalism}. Note that the saddle point equation is nothing more than the variation principle stated in Eq.~(\ref{variationPrinciple2}). Thus, solving the saddle point equations is equivalent to determining the total resistance $R_r \left( \left\{ t_i \right\} \right)$ of a diagram, and the saddle point evaluation of (\ref{toEvaluate}) yields
\begin{eqnarray}
\exp \left[ R_r \left(  \left\{ t_i \right\} \right) \rho \, w_r \Lambda_r \left( \vec{\lambda} \right) \right] \ , 
\end{eqnarray}
where we have omitted once more multiplicative factors which go to one for $D \to0$. A completion of squares in the momenta renders the momentum integrations, which remain to be done to compute the diagrams, straightforward. Equally well we can use the saddle point method which is exact here since the momentum dependence is purely quadratic. After an expansion for small $w_r \Lambda_r \left( \vec{\lambda} \right)$ all diagrammatic contributions are of the form
\begin{eqnarray}
\label{expansionOfDiagrams}
I \left( {\rm{\bf p}}^2 , t, \Lambda_r \left( \vec{\lambda} \right) \right) &=& I_P \left( {\rm{\bf p}}^2 , t \right) + I_W 
\left( {\rm{\bf p}}^2 , t \right) \rho \, w_r  \Lambda_r \left( \vec{\lambda} \right) + \cdots
\nonumber \\
&=& \int_0^\infty \prod_i dt_i \left[ 1 + R_r \left(  \left\{ t_i \right\} 
\right) \rho \, w_r \Lambda_r \left( \vec{\lambda} \right) + \cdots \right] D \left( {\rm{\bf p}}^2,t ; 
\left\{ t_i \right\} \right) \ .
\end{eqnarray}
$D \left( {\rm{\bf p}}^2, t ; \left\{ t_i \right\} \right)$ is a typical integrand as known from the field theory of DP\cite{cardy_sugar_80,janssen_81,janssen_2000}.

\subsection{Renormalization and scaling}
\label{renormalizationAndScaling}
We proceed with standard techniques of renormalized field theory\cite{amit_zinn-justin}. The ultraviolet divergences occurring in the diagrams can be absorbed by dimensional regularization. We employ the renormalization scheme
\begin{mathletters}
\label{renorScheme}
\begin{eqnarray}
s \to {\mathaccent"7017 s} = Z^{1/2} s \ ,&\quad&
\rho \to {\mathaccent"7017 \rho} = Z^{-1} Z_{\rho} \rho \ ,
\\
\tau \to {\mathaccent"7017 \tau} = Z^{-1}_\rho Z_{\tau} \tau \ , &\quad&
w_r \to {\mathaccent"7017 w_r} = Z^{-1}_\rho Z_{w_r} w_r \ , 
\\
\overline{g} \to \mathaccent"7017{\overline{g}} &=& Z^{-1/2} Z^{-1}_\rho Z_u^{1/2} G_\epsilon^{-1/2} u^{1/2} 
\mu^{\epsilon /2} \ ,
\end{eqnarray}
\end{mathletters}
where $\epsilon = 4-d_\perp$ and $\mu$ is the usual inverse length scale. The factor $G_\epsilon = (4\pi )^{-d_\perp /2}\Gamma (1 + \epsilon /2)$, with $\Gamma$ denoting the Gamma function, is introduced for convenience. $Z$, $Z_\tau$, $Z_\rho$, and $Z_u $ are the usual DP $Z$ factors known to second order in $\epsilon$\cite{janssen_81,janssen_2000,stenull_janssen_directedResistance_2000}. In Ref.~\cite{stenull_janssen_directedResistance_2000} we determined $Z_w = Z_{w_1}$ to second order in $\epsilon$. Here, we calculate $Z_{w_r}$ for arbitrary $r$ to order $\epsilon$. This calculation is straightforward because we can determine the total resistance of the one-loop diagrams by using simple rules. For example, two nonlinear resistors with resistances $t_1$ and $t_2$ added in series have a total resistance $R_r$ given by
\begin{eqnarray}
\label{addrule1}
R_r (t_1 , t_2 ) = t_1 +  t_2 \ ,
\end{eqnarray}
whereas two such resistors in parallel give
\begin{eqnarray}
\label{addrule2}
R_r (t_1 , t_2 )^{-s} = t_1^{-s} +  t_2^{-s} \ .
\end{eqnarray} 
By exploiting Eq.~(\ref{addrule2}) find
\begin{eqnarray}
Z_{w_r} = 1 + \frac{u}{2 \, \epsilon} \left[ 1 - \frac{1}{2^{r+1}} \right] + {\sl O} \left( u^2 \right) \ .
\end{eqnarray}
Calculating $Z_{w_r}$ for general $r$ to higher loop orders appears to be beyond possibility. The reason is, that conducting diagrams like C appear. The total resistance of these diagrams cannot be determined by using simple rules like Eqs.~(\ref{addrule1}) and (\ref{addrule2}). Instead, one has to solve the set of nonlinear circuit equations which is hardly feasible in closed form.

Now we set up in a standard fashion the renormalization group equation for our problem. The unrenormalized theory has to be independent of the length scale $\mu^{-1}$ introduced by renormalization. In particular, the unrenormalized connected $N$ point correlation functions must be independent of $\mu$, i.e.,  
\begin{eqnarray}
\label{independence}
\mu \frac{\partial}{\partial \mu} {\mathaccent"7017 G}_N \left( \left\{ {\rm{\bf x}}_\perp , {\mathaccent"7017 \rho} t , {\mathaccent"7017 w}_r \Lambda_r \left( \vec{\lambda} \right) \right\} ; {\mathaccent"7017 \tau}, \mathaccent"7017{\overline{g}} \right) = 0
\end{eqnarray}
for all $N$. Eq.~(\ref{independence}) translates via the Wilson functions
\begin{mathletters}
\begin{eqnarray}
\label{wilson}
\beta \left( u \right) = \mu \frac{\partial u}{\partial \mu} \bigg|_0 \ ,
&\quad& 
\kappa \left( u \right) = \mu \frac{\partial
\ln \tau}{\partial \mu}  \bigg|_0 \ ,
 \\
\zeta_{r} \left( u \right) = \mu \frac{\partial \ln w_r}{\partial \mu}  \bigg|_0 \ ,
&\quad&
\zeta_\rho \left( u \right) = \mu \frac{\partial \ln \rho}{\partial \mu}  \bigg|_0 \ ,
\\
\gamma_{...} \left( u \right) &=& \mu \frac{\partial }{\partial \mu} \ln Z_{...}  \bigg|_0 \ ,
\end{eqnarray}
\end{mathletters}
where the bare quantities are kept fix while taking the derivatives, into the Gell-Mann-Low renormalization group equation
\begin{eqnarray}
\label{rge}
\lefteqn{ 
\left[ \mu \frac{\partial }{\partial \mu} + \beta \frac{\partial }{\partial u} + 
\tau \kappa \frac{\partial }{\partial \tau} + w_r \zeta_{r} \frac{\partial }{\partial w_r} + \rho \zeta_\rho \frac{\partial }{\partial \rho} +
\frac{N}{2} \gamma \right] 
}
\nonumber \\
&& \times \, G_N \left( \left\{ {\rm{\bf x}}_\perp , \rho t  ,w_r \Lambda_r \left( \vec{\lambda} \right) \right\} ; \tau, u, \mu \right) = 0 \ .
\end{eqnarray}
The particular form of the Wilson functions can be extracted from the renormalization scheme and the $Z$ factors.

The critical behavior of the correlation functions is determined by the infrared stable fixed point solutions of Eq.~(\ref{rge}). This fixed point $u^\ast$ is readily extracted from the condition $\beta \left( u^\ast \right) = 0$. Then Eq.~(\ref{rge}) is solved at $u^\ast$ by the method of characteristics which gives
\begin{eqnarray}
\label{SolOfRgg}
\lefteqn{ G_N \left( \left\{ {\rm{\bf x}}_\perp , \rho t ,w_r \Lambda_r \left( \vec{\lambda} \right) \right\} ; \tau, u, \mu \right) } 
\nonumber \\
&& = \, 
l^{\gamma^\ast N/2} G_N \left( \left\{ l{\rm{\bf x}}_\perp ,l^{\zeta_\rho^\ast}\rho t , l^{\zeta_{r}^\ast}w_r \Lambda_r \left( \vec{\lambda} \right)\right\} ; l^{\kappa^\ast}\tau , u^\ast, l \mu \right) \ , 
\end{eqnarray}
where $\gamma^\ast = \gamma \left( u^\ast \right)$, $\kappa^\ast = \kappa \left( u^\ast \right)$, $\zeta_\rho^\ast = \zeta_\rho \left( u^\ast \right)$ , and $\zeta_r^\ast = \zeta_r \left( u^\ast \right)$. To analyze the scaling behavior of the correlation functions completely, the solution~(\ref{SolOfRgg}) has to be supplemented by a dimensional analysis:
\begin{eqnarray}
\label{dimAna}
\lefteqn{ G_N \left( \left\{ {\rm{\bf x}}_\perp ,\rho t ,w_r \Lambda_r \left( \vec{\lambda} \right) \right\} ; \tau, u, \mu \right) }
\nonumber \\
&& = \,  
\mu^{d_\perp N/2} G_N \left( \left\{ \mu {\rm{\bf x}}_\perp , \mu^2 \rho t , \mu^{-2}w_r \Lambda_r \left( \vec{\lambda} \right) \right\} ; \mu^{-2}\tau , u, 1 \right) \ . 
\end{eqnarray}
Equation~(\ref{SolOfRgg}) in conjunction with Eq.~(\ref{dimAna}) now gives
\begin{eqnarray}
\label{scaling}
\lefteqn{ G_N \left( \left\{ {\rm{\bf x}}_\perp , \rho t ,w_r \Lambda_r \left( \vec{\lambda} \right) \right\} ; \tau, u, \mu \right) }
\nonumber \\
&& = \, 
l^{(d_\perp +\eta)N/2} G_N \left( \left\{ l{\rm{\bf x}}_\perp , l^z \rho t , l^{-\phi_r /\nu_\perp} w_r \Lambda_r \left( \vec{\lambda} \right)  \right\} ; 
l^{-1/\nu_\perp}\tau , u^\ast, \mu \right) \ .
\end{eqnarray}
$\eta = \gamma^\ast$, $z = 2 + \zeta_\rho^\ast$, and $\nu_\perp = \frac{1}{2-\kappa^\ast }$
are the well known the critical exponents for DP which have been calculated previously to second order in $\epsilon$\cite{janssen_81,janssen_2000,stenull_janssen_directedResistance_2000}. These DP exponents, however, are not sufficient to specify the critical behavior of the RDN correlation functions completely. In Eq.~(\ref{scaling}) we introduced the additional nonlinear resistance exponent
\begin{eqnarray}
\label{resPhi}
\phi_r &=& \nu_\perp \left( 2 - \zeta_r^\ast \right) = 1 + \frac{\epsilon}{3 \cdot 2^{r+2}} + {\sl O} \left( \epsilon^2 \right) \ .
\end{eqnarray}
Note that $\phi = \phi_1$ is in conformity to order $\epsilon$ with our result for the resistance exponent for the usual ``ohmic'' RDN, i.e., Eq.~(\ref{resPhi}) satisfies an important consistency check.

Since we are primarily interested in the critical behavior of the average two-port resistance, we now take a closer look at the two point correlation function $G=G_2$. Equation~(\ref{scaling}) implies for $G$ at $\tau = 0$ that
\begin{eqnarray}
\label{scaleRel}
G \left( |{\rm{\bf x}}_\perp-{\rm{\bf x}}_\perp^\prime |, t-t^\prime , w_r \Lambda_r \left( \vec{\lambda} \right) \right) = 
l^{d_\perp +\eta} G \left( l |{\rm{\bf x}}_\perp-{\rm{\bf x}}_\perp^\prime|,l^z \left( t - t^\prime \right) , l^{-\phi_r /\nu_\perp} w_r \Lambda_r \left( \vec{\lambda} \right) \right) \ ,
\end{eqnarray}
where we dropped several arguments for notational simplicity. In the following we set ${\rm{\bf x}}_\perp^\prime = {\rm{\bf 0}}$ and $t^\prime =0$, once more for the sake of simplicity. The choice $l = |{\rm{\bf x}}_\perp|^{-1}$ and a Taylor expansion of the right hand side of 
Eq.~(\ref{scaleRel}) lead to 
\begin{eqnarray}
\label{scaleX}
G \left( |{\rm{\bf x}}_\perp|, t , w_r \Lambda_r \left( \vec{\lambda} \right) \right) 
&=& 
\left| {\rm{\bf x}}_\perp \right|^{1-d-\eta} f_1 \left(  \frac{t}{\left| {\rm{\bf x}}_\perp \right|^z} \right)
\nonumber \\
&& \times \, 
 \bigg\{ 1 + w_r \Lambda_r \left( \vec{\lambda} \right) \left| {\rm{\bf x}}_\perp \right|^{\phi_r /\nu_\perp} f_{w,1} \left(  \frac{t}{\left| {\rm{\bf x}}_\perp \right|^z} \right) + \cdots \bigg\} \ ,
\end{eqnarray}
where the $f$'s are scaling functions. Equally well we can choose $l=t^{-1/z}$ which then leads to
\begin{eqnarray}
\label{scaleT}
 G \left( |{\rm{\bf x}}_\perp|, t , w_r \Lambda_r \left( \vec{\lambda} \right) \right)  
&=&  
t^{(1-d-\eta)/z} f_2 \left(  \frac{\left| {\rm{\bf x}}_\perp \right|^z}{t} \right) \nonumber \\
&& \times \, 
\bigg\{ 1 + w_r \Lambda_r \left( \vec{\lambda} \right) \, t^{\phi_r /\nu_\parallel} f_{w,2} \left(  \frac{\left| {\rm{\bf x}}_\perp \right|}{t} \right) + \cdots \bigg\} \ ,
\end{eqnarray}
with other scaling functions $f_2$ and $f_{w,2}$ and where $\nu_\parallel = \nu_\perp z$.
 
Now we can extract the critical behavior of $M_{R_r}$. For measurements in the distinguished direction we straightforwardly exploit Eq.~(\ref{scaleT}) via Eq.~(\ref{shit}) and find that
\begin{eqnarray}
\label{keineLustMehr} 
M_{R_r} \sim t^{\phi_r /\nu_\parallel} \ .
\end{eqnarray}
For all other directions we determine $M_{R_r}$ from Eq.~(\ref{scaleX}). With help of Eq.~(\ref{shit}) we find that
\begin{eqnarray}
\label{brauchNoch} 
M_{R_r} \sim \left| {\rm{\bf x}}_\perp \right|^{\phi_r /\nu_\perp} f_{w,1} \left(  \frac{t}{\left| {\rm{\bf x}}_\perp \right|^z} \right) \ .
\end{eqnarray}
Here it is convenient to choose a common length scale $L$ and to express both, $\left| {\rm{\bf x}}_\perp \right|$ and $t$, in terms of it: $\left| {\rm{\bf x}}_\perp \right| \sim L$ and $t \sim L^z$. This choice guarantees that the scaling functions $f_{w,1}$ is constant and Eq.~(\ref{brauchNoch}) simplifies to
\begin{eqnarray} 
M_{R_r} \sim L^{\phi_r /\nu_\perp} \ .
\end{eqnarray}

\section{Fractal dimensions of DP clusters}
\label{fracDim}
In this section we calculate $\phi_r$ for $r \to 0^+$, $r \to \infty$, and $r \to -1^+$ to two-loop order. This provides us with the fractal dimension of the backbone, the red bonds, and the chemical length respectively.

\subsection{Red bonds}
\label{redBonds}
For self-affine objects the notion of fractal dimension is less straightforward than for self-similar objects. To determine, for example, the fractal dimension of the red bonds in DP one considers a $d-1$-dimensional hyper-plane with an orientation perpendicular to $x_\parallel$. The cut through the red bonds is a self similar object with the fractal dimension
\begin{eqnarray}
\label{cutDim}
d^{\mbox{\scriptsize{cut}}}_{\mbox{\scriptsize{red}}} = d_{\mbox{\scriptsize{red}}}- 1 \ ,
\end{eqnarray}
where $d_{\mbox{\scriptsize{red}}}$ is the local fractal dimension\cite{kertez_vicsek_94} of the red bonds. By virtue of Eqs.~(\ref{massRed}) and (\ref{brauchNoch}) the mass of the red bonds scales as
\begin{eqnarray}
\label{RedMassScaling}
M_{\mbox{\scriptsize{red}}} = \left| {\rm{\bf x}}_\perp \right|^{\phi_{\infty} /\nu_\perp} f_{w,1} \left(  \frac{x_\parallel}{\left| {\rm{\bf x}}_\perp \right|^z} \right) \ .
\end{eqnarray}
Accordingly the mass of the cut scales like
\begin{eqnarray}
\label{cutScaling}
M^{\mbox{\scriptsize{cut}}}_{\mbox{\scriptsize{red}}} = \left| {\rm{\bf x}}_\perp \right|^{\phi_{-1} /\nu_\perp} x_\parallel^{-1} f_{w,1} \left(  \frac{x_\parallel}{\left| 
{\rm{\bf x}}_\perp \right|^z} \right) \ .
\end{eqnarray}
By choosing once more $\left| {\rm{\bf x}}_\perp \right| \sim L$ and $x_\parallel \sim L^z$ we find that 
\begin{eqnarray}
M^{\mbox{\scriptsize{cut}}}_{\mbox{\scriptsize{red}}} \sim L^{\phi_{\infty} /\nu_\perp - z} \ . 
\end{eqnarray}
This leads via Eq.~(\ref{cutDim}) to
\begin{eqnarray}
\label{dRedFoermelchen}
d_{\mbox{\scriptsize{red}}} = 1 + \phi_{\infty} /\nu_\perp - z \ .
\end{eqnarray}

It remains to compute $\phi_{\infty}$. To do so we take direct advantage of our view of the Feynman diagrams as being resistor networks themselves. As argued in Sec.~\ref{clusterProperties}, blobs do not contribute to the total resistance for $r \to \infty$. In analogy only singly connected conducting propagators contribute to the total resistance of a diagram, i.e,
\begin{eqnarray}
R_\infty \left( \left\{ t_i \right\} \right) =  \sum_i^{\mbox{\scriptsize singly}} t_i\ ,
\end{eqnarray}
with the sum being taken exclusively over singly connected conducting propagators. The contribution of a diagram to the renormalization factor $Z_{w_\infty}$ takes the form 
\begin{eqnarray}
\label{expansionOfDiagrams_r_infty}
I_W \left( {\rm{\bf p}}^2 , t\right) = \int_0^\infty \prod_j dt_j \,  D \left( {\rm{\bf p}}^2, t ; \left\{ t_j \right\} \right)  \sum_i^{\mbox{\scriptsize singly}} t_i \ .
\end{eqnarray}
Note that a factor $t_i$ in Eq.~(\ref{expansionOfDiagrams_r_infty}) corresponds to the insertion (cf.~Ref.\cite{amit_zinn-justin}) of $\frac{1}{2} s^2$ into the $i$th edge of the diagram. Thus, we can generate $I_W \left( {\rm{\bf p}}^2 , t \right)$ for a given conducting diagram by inserting $\frac{1}{2} s^2$ into its singly connected conducting propagators. This procedure is carried out up to two-loop order, i.e., every conducting propagator in Fig.~1 that does not belong to a closed loop gets an insertion. The resulting diagrams are displayed in Fig.~2.

At this point it is instructive to consider the contributions of the diagrams listed in Fig.~1 to $Z_\tau$. These can be generated by inserting $\frac{1}{2} s^2$ in conducting as well as in insulating propagators. Again, one obtains the diagrams depicted in Fig.~2 with the same fore-factors. Consequently, $Z_{w_\infty}$ and $Z_\tau$ are identical at least up to two-loop order. The same goes for the corresponding Wilson functions $\zeta_\infty$ and $\kappa$. From the definition of $\phi_r$ it 
follows that
\begin{eqnarray}
\label{resultForPhiInfty}
\phi_\infty = \frac{2-\zeta_\infty^\ast}{2-\kappa^\ast} = 1 + {\sl O} \left( \epsilon^3  \right) \ .
\end{eqnarray}
Upon inserting Eq.~(\ref{resultForPhiInfty}) into Eq.~(\ref{dRedFoermelchen}) we obtain the prime result of Sec.~\ref{redBonds},
\begin{eqnarray}
\label{dRedResult}
d_{\mbox{{\scriptsize red}}} = 1 + 1 /\nu_\perp - z \ ,
\end{eqnarray}
holding at least to second order in $\epsilon$. By substituting the results of Ref.~\cite{janssen_81,janssen_2000} in Eq.~(\ref{dRedResult}) we obtain the following $\epsilon$-expansion for $d_{\mbox{{\scriptsize red}}}$:
\begin{eqnarray}
\label{dRedExpansion}
d_{\mbox{{\scriptsize red}}} = 1 - \frac{\epsilon}{6} \left\{1 + \left[ \frac{73}{288} - \frac{55}{144} \ln \left( \frac{4}{3} \right) \right] \, \epsilon  \right\} + {\sl O} \left( \epsilon^3  \right) \ .
\end{eqnarray}

Note that the scaling relation (\ref{dRedResult}) holds rigorously and its validity is not restricted to second order in the $\epsilon$-expansion. Coniglio\cite{coniglio_81&82} proved for IP that the mass of the red bonds scales as
\begin{eqnarray}
\label{iso}
M_{\mbox{{\scriptsize red}}} \left( {\rm{\bf 0}},{\rm{\bf x}} \right) \sim \left| {\rm{\bf x}} \right|^{1/\nu} \ .
\end{eqnarray}
For IP this is equivalent to saying that the fractal dimension of the red bonds is $d_{\mbox{{\scriptsize red}}} = 1/\nu$. Since Coniglio's arguments do not rely on the isotropy of the system, they can be adapted to apply to DP. For the DP problem Eq.~(\ref{iso}) has to be modified to
\begin{eqnarray}
M_{\mbox{{\scriptsize red}}} \left( {\rm{\bf 0}},{\rm{\bf x}} \right) = \left| {\rm{\bf x}}_\perp \right|^{1 /\nu_\perp} f_{w,1} \left(  \frac{x_\parallel}{\left| {\rm{\bf x}}_\perp \right|^z} \right) \ .
\end{eqnarray}
This in turn leads again to Eq.~(\ref{dRedResult}).

\subsection{Chemical length}
\label{chemicalLenght}
Next we address the fractal dimension of the chemical length. Equation~(\ref{dMinRel}) in conjunction with Eq.~(\ref{brauchNoch}) provides us with
\begin{eqnarray}
\label{minMassScaling}
M_{\mbox{{\scriptsize min}}} = \left| {\rm{\bf x}}_\perp \right|^{\phi_{0} /\nu_\perp} f_{w,1} \left(  \frac{x_\parallel}{\left| {\rm{\bf x}}_\perp \right|^z} \right) \ .
\end{eqnarray}
By applying the same reasoning as in Sec.~\ref{backbone} we learn that the local fractal dimension of the chemical length is given by
\begin{eqnarray}
\label{dMinFoermelchen}
d_{\mbox{{\scriptsize min}}} = 1 + \phi_{0} /\nu_\perp - z \ .
\end{eqnarray}

In order to calculate $\phi_{0}$ we determine the shortest self-avoiding path of conducting propagators connecting the external legs of a diagram. Due to the dynamic structure all of these paths for a given diagram have the same total resistance that is nothing more than the total ``time'' between the external legs. Hence, we can choose any self-avoiding path connecting the external legs. We work with the diagrammatic expansion depicted in Fig.~3.

Minimal subtraction provides us with the renormalization factor 
\begin{eqnarray}
\label{renFuck}
Z_0 = 1 + \frac{u}{4 \, \epsilon} + \frac{u^2}{32 \, \epsilon} \left[ \frac{7}{\epsilon} - 3 + \frac{9}{2} \ln \left( \frac{4}{3} \right) \right] + {\sl O} \left( u^3 \right) \ .
\end{eqnarray}
Note that $Z_0$ as stated in Eq.~(\ref{renFuck}) is identical to the field renormalization $Z$ given to two-loop order in Refs.~\cite{janssen_81,janssen_2000,stenull_janssen_directedResistance_2000}. By virtue of the renormalization scheme (\ref{renorScheme}) we deduce that
\begin{eqnarray}
\label{trivialRen}
\mathaccent"7017{\rho} \, \mathaccent"7017{w}_0 = \rho \, w_0 \ ,
\end{eqnarray}
at least to second order in $u$. Equation~(\ref{trivialRen}) leads via
\begin{eqnarray}
\mu \, \frac{\partial \ln \left( \rho \, w_0 \right) }{\partial \mu}  \bigg|_0 = 0
\end{eqnarray}
to
\begin{eqnarray}
\zeta_\rho +  \zeta_0 = 0 \ .
\end{eqnarray}
From the definitions of $\phi_r$ and $z$ it follows immediately that
\begin{eqnarray}
\phi_0 = \nu_\perp \, z + {\sl O} \left( \epsilon^3 \right) \ .
\end{eqnarray}
In conjunction with Eq~(\ref{dMinRel}) this leads finally to
\begin{eqnarray}
\label{resDMin}
d_{\mbox{{\scriptsize min}}} = 1 + {\sl O} \left( \epsilon^3 \right) \ . 
\end{eqnarray}
This result is intuitively plausible because the chemical distance in DP is basically equivalent to the ``time'' $t$.

\subsection{Backbone}
\label{backbone}
We conclude Sec.~\ref{fracDim} by studying the backbone dimension. By virtue of Eqs.~(\ref{massBB}) and (\ref{brauchNoch}) the mass of the backbone scales as
\begin{eqnarray}
\label{massScaling}
M_B = \left| {\rm{\bf x}}_\perp \right|^{\phi_{-1} /\nu_\perp} f_{w,1} \left(  \frac{x_\parallel}{\left| {\rm{\bf x}}_\perp \right|^z} \right) \ .
\end{eqnarray}
Accordingly the local fractal dimension of the backbone is given by
\begin{eqnarray}
\label{DBfoermelchen}
D_B = 1 + \phi_{-1} /\nu_\perp - z \ .
\end{eqnarray}

It remains to compute $\phi_{-1}$. Once more we exploit our real-world interpretation. As argued in Sec.~\ref{clusterProperties}, the resistance of the backbone between two sites $x$ and $x^\prime$ is given by
\begin{eqnarray}
R_{-1,+} (x ,x^\prime) =\sum_{\underline{b}} \rho_{\underline{b}} \left( I_{\underline{b}}  \right) \ ,
\end{eqnarray}
with the sum running over all current carrying bonds of the underlying cluster. In analogy, the resistance of a conducting Feynman diagram is given by
\begin{eqnarray}
R_{-1} \left( \left\{ t_i \right\} \right) =  \sum_i^{\mbox{\scriptsize cond}} t_i\ ,
\end{eqnarray}
where the sum is extending over all conducting propagators of the diagram. This tells us that the contribution of a diagram to $Z_{w_{-1}}$ takes the form 
\begin{eqnarray}
\label{expansionOfDiagrams_r_-1}
I_W \left( {\rm{\bf p}}^2 , t \right) = \int_0^\infty \prod_j dt_j \,  
D \left( {\rm{\bf p}}^2, t , \left\{ t_j \right\} \right) \sum_i^{\mbox{\scriptsize cond}} t_i \ .
\end{eqnarray}
Hence we can generate $I_W \left( {\rm{\bf p}}^2 , t \right)$ of any conducting diagram by inserting $\frac{1}{2} s^2$ into its conducting propagators. All conducting propagators in Fig.~1 get such an insertion. Then it is a matter of simple counting to see that the individual contributions cancel each other. Thus, we find
\begin{eqnarray}
Z_{w_{-1}} = 1 + {\sl O} \left( \epsilon^3  \right) \ .
\end{eqnarray}
As a consequence we obtain
\begin{eqnarray}
\label{SkalenRel}
\lim_{r \to -1^+} \phi_r /\nu_\perp = z - \eta \ ,
\end{eqnarray}
at least to second order in $\epsilon$. Equation~(\ref{SkalenRel}) leads by virtue of Eq.~(\ref{DBfoermelchen}) to
\begin{eqnarray}
\label{ergebnisDB}
D_B = 1- \eta = d - \frac{2\beta}{\nu_\perp} \ ,
\end{eqnarray}
where  $\beta = \nu_\perp (d-1+\eta )/2$ is the DP order parameter exponent known to second order in $\epsilon$\cite{janssen_81,janssen_2000}. From the scaling relation Eq.~(\ref{ergebnisDB}), which is the main result of this section, the $\epsilon$-expansion of $D_B$ is readily obtained by inserting the $\epsilon$-expansion for $\eta$\cite{janssen_81,janssen_2000}:
\begin{eqnarray}
\label{expansionDB}
D_B = 1 + \frac{\epsilon}{6} \left\{ 1 + \left[ \frac{25}{288} + \frac{161}{144}\ln \left( \frac{4}{3} \right) \right] \epsilon \right\} + {\sl O} \left( \epsilon^3  \right) \ .
\end{eqnarray}

Equation~(\ref{ergebnisDB}) is in agreement with scaling arguments\cite{hede&co_91} yielding that the fractal dimension of the transverse cut through a DP cluster with local dimension $d_f$ is $d_f -1 = d-1 -\beta / \nu_\perp$. The analogous cut through the backbone can be viewed as the intersection of the cut through the cluster and the clusters backward oriented pendant\cite{red_83,arora&co_83}. Hence, the codimension of the backbone cut is twice the codimension $\beta / \nu_\perp$ of the cluster cut, which leads again to Eq.~(\ref{ergebnisDB}).

It is interesting to compare the $\epsilon$-expansion result to numerical estimates. We are not aware, however, of any simulations in which $D_B$ itself was determined. Huber {\em et al}.~\cite{sneppen&co_95} presented numerical results for the scaling exponent of the backbone mass when measured in the longitudinal direction. In the following we call this exponent $\tilde{D}_B$. Formally one can define $\tilde{D}_B$ via $M_B \sim t^{\tilde{D}_B}$. From Eqs.~(\ref{massBB}), (\ref{keineLustMehr}), and (\ref{SkalenRel}) it follows that
\begin{eqnarray}
\tilde{D}_B = 1 + \frac{2\beta}{\nu_\parallel} + \frac{d-1}{z}  \ ,
\end{eqnarray}
at least to second order in $\epsilon$. Crudely evaluating the corresponding $\epsilon$-expansion of $\tilde{D}_B$ for small spatial dimensions leads to poor quantitative predictions. Therefore it is appropriate to improve the $\epsilon$-expansion by incorporating rigorously known features. We carry out a rational approximation which takes into account that obviously $\tilde{D}_B (d=1) =1$. Practically this is done by adding an appropriate third order term. By this procedure we obtain the interpolation formula
\begin{eqnarray}
\label{DbAppr}
\tilde{D}_B \approx 1 + \left( 1 - \frac{\epsilon}{4} \right) \left( 0.0833 \, \epsilon + 0.0583 \, \epsilon^2 \right) \ .
\end{eqnarray}
Evaluation in two dimensions leads to $\tilde{D}_B (d=2) = 1.2 \pm 0.15$, where the error is based on a subjective estimate. This result is, within the errors, in agreement with the numerical result\cite{sneppen&co_95} $\tilde{D}_B (d=2) = 1.30 \pm 0.03$. Our result though appears to be somewhat small.

\section{conclusions}
\label{concls}
In this paper we studied a nonlinear version of resistor diode percolation where Ohm's law is generalized to $V \sim I^r$. We investigated the critical behavior of the average two-port resistance $M_{R_r}$ at the transition from the non percolating to the directed percolating phase. By employing our real-world interpretation of Feynman diagrams we calculated the resistance exponent $\phi_r$ for arbitrary $r$ to one-loop order. To our knowledge this is the first time that $\phi_r$ has been determined for the RDN while $\phi_r$ is known for RRN, also to one-loop order, since the 1980's\cite{harris_87}. Extending either of these results to higher loop orders seems to be beyond possibility because not all conducting diagrams appearing at higher loop orders can be assembled by simply adding resistors in parallel and in series. For these diagrams one has to solve the set of nonlinear Kirchhoff's equations to obtain their total resistance. In closed form, however, this is hardly feasible.

The relation of $M_{R_r}$ to the mass of the red bonds, the chemical length, and the backbone, respectively, provided us with alternative means to extract the fractal dimensions of these substructures of DP clusters. By computing $\phi_r$ for $r \to \infty$, $r \to 0^+$, and $r \to -1^+$ we determined $d_{\mbox{{\scriptsize red}}}$, $d_{\mbox{{\scriptsize min}}}$, and $D_B$ to two-loop order.

The fractality in DP and IP is qualitatively different. DP clusters are self affine rather than self similar objects. Hence, the notion of fractal dimensions in more subtle for DP than for IP. Moreover, DP has a Markovian character which is evident in the dynamic interpretation. This Markovian character provides for scaling relations which do not have an analog in IP. The DP backbone dimension for example can be expressed entirely in terms of the usual (purely geometric) critical exponents of the DP universality class, $D_B = d - 2\beta /\nu_\perp$. Within the renormalization group framework such a scaling relation is typically associated with a Ward identity. The fact that $w_{-1}$ renormalizes trivially to two-loop order is reminiscent of this Ward identity. It is an interesting issue for future work to identify the Ward identity and its underlying symmetry. Another consequence of the Markovian character of DP is that the fractal dimension of the chemical length is identical to one. This is intuitively plausible since the shortest longitudinal path through a DP cluster corresponds to the time in the dynamical interpretation.

The fractal dimension of the red bonds in DP obeys the scaling relation $d_{\mbox{{\scriptsize red}}} = 1 + 1/\nu_\perp - z$ similar to the $d_{\mbox{{\scriptsize red}}} =  1/\nu$ for IP. The Ward identities corresponding to either of these scaling relations are not known to date. Again, this leaves interesting and challenging opportunities for future studies. 

\acknowledgements
We acknowledge support by the Sonderforschungsbereich 237 ``Unordnung und gro{\ss}e Fluktuationen'' of the Deutsche Forschungsgemeinschaft.  

%appendices

%references

%
%%%%%%
%
%figures
\newpage
\begin{figure}[h]
\epsfxsize=10.0cm
\begin{center}
\epsffile{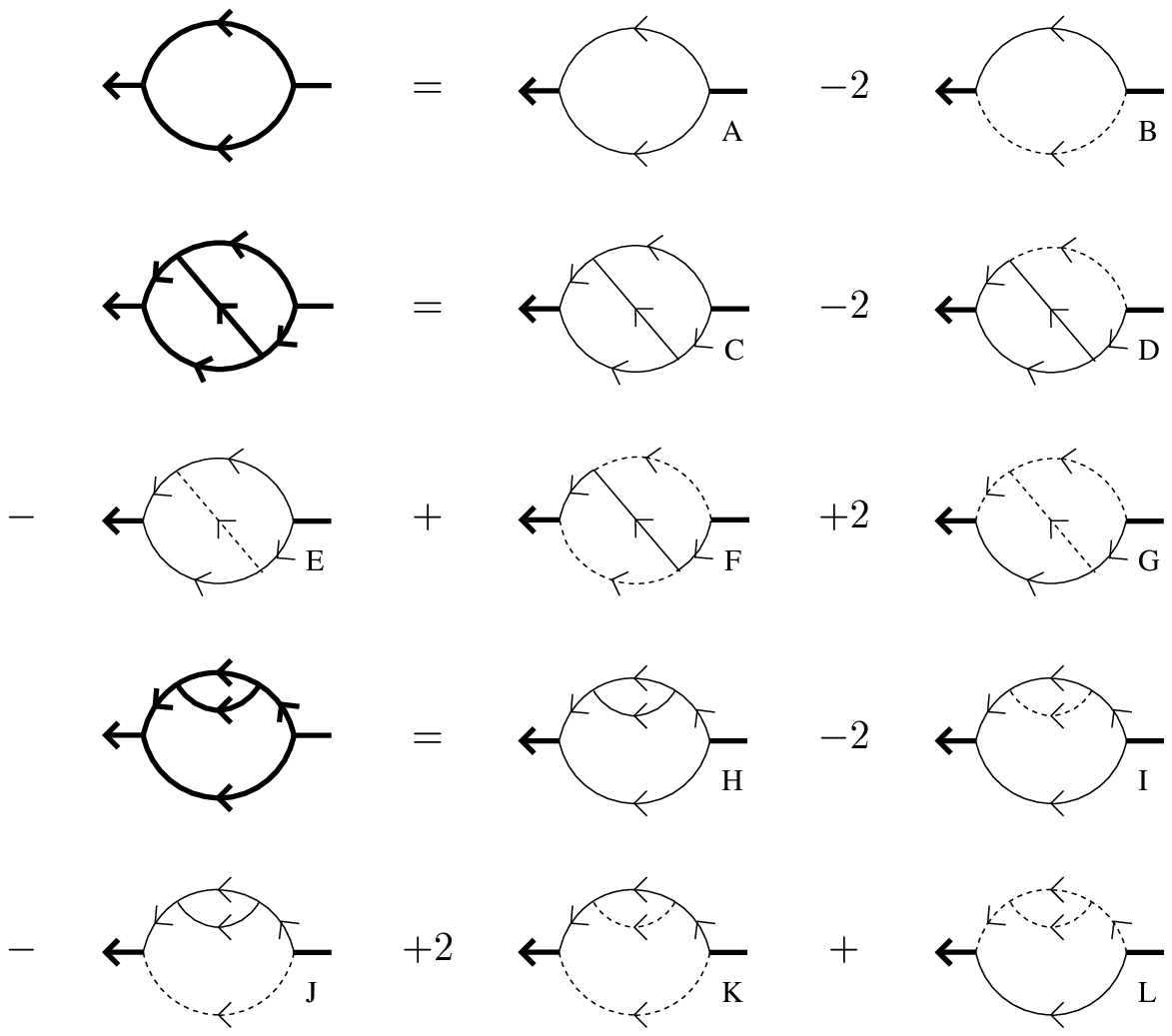}
\end{center}
\caption[]{Decomposition of the primary two leg diagrams (bold) into conducting diagrams composed of conducting (light) and insulating (dashed) propagators to two-loop order. It is important to realize that the conducting diagrams inherit their combinatorial factor from their bold diagram. For example, the diagrams A and B introduced below have to be calculated with the same combinatorial factor $\frac{1}{2}$.}
\end{figure}
%
%\newpage
\begin{figure}[h]
\epsfxsize=10.0cm
\begin{center}
\epsffile{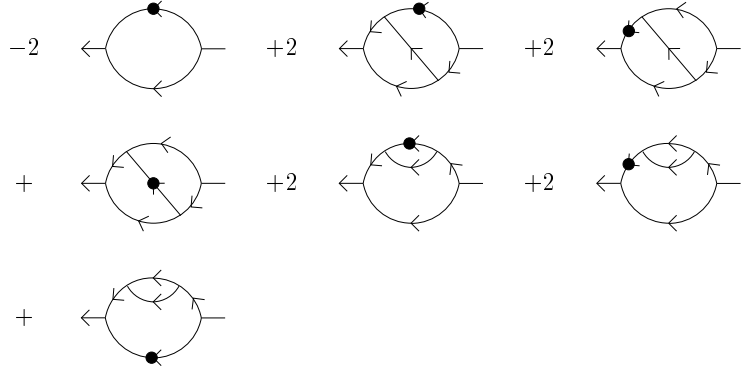}
\end{center}
\caption[]{Diagrammatic expansion in the limit $r \to \infty$. The listed diagrams including their fore-factors can be obtained from the conducting diagrams shown in Fig.~1 in two different ways: first, by inserting $\frac{1}{2} s^2$ into all singly connected conducting propagators and second, by inserting $\frac{1}{2} s^2$ into every conducting and insulating propagator. As a consequence, the renormalization factors $Z_{w_\infty}$ and $Z_\tau$ are identical. The lines stand for conducting propagators evaluated at zero currents, the solid dots for $\frac{1}{2} s^2$-insertions.}
\end{figure}
\newpage
\begin{figure}[h]
\epsfxsize=10.0cm
\begin{center}
\epsffile{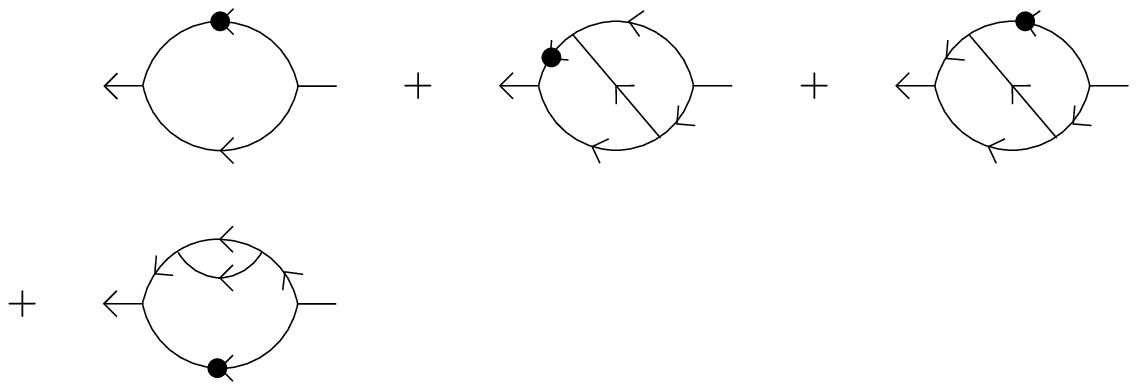}
\end{center}
\caption[]{Diagrammatic expansion in the limit $r \to 0^+$. The meaning of the symbols is the same as in Fig.~2.}
\end{figure}

\begin{references}
\bibitem{bunde_havlin_91_etc} 
For a review on percolation see, e.g., A. Bunde and S. Havlin, {\em Fractals and Disordered Systems} (Springer, Berlin, 1991);  D. Stauffer and A. Aharony, {\em Introduction to Percolation Theory} (Taylor{\&}Francis, London, 1992); B. D. Hughes, {\em Random Walks and Random Environments} (Clarendon, Oxford, 1995).
\bibitem{hinrichsen_2000}
For a review on DP see, e.g., H. Hinrichsen, Adv. in Phys. {\bf 49}, 815 (2000).
\bibitem{vanLien_shklovskii_81}
N. Van Lien and B. I. Shklovskii, Solid State Commun.\ {\bf 38}, 99 (1981).
\bibitem{kertez_vicsek_80}
J. Kert\'{e}sz and T. Vicsek, J.\ Phys.\ C {\bf 13}, L343 (1980).
\bibitem{depinning}
See, e.g., A.-L. Barabasi {\em et al}., in {\em Surface Disordering: Growth, Roughening, and Phase Transitions}, edited by R. Jullien {\em et al}. (Nova Science, New York, 1992); S. Havlin {\em et al}., in {\em Growth Patterns}, edited by J. M. Garcia-Ruiz {\em et al}. (Plenum, New York, 1993).
\bibitem{murray_88}
See, e.g., J. D. Murray, {\em Mathematical Biology} (Springer, Berlin, 1988).
\bibitem{grassberger_85}
P. Grassberger, J.\ Phys.\ A {\bf 18}, L215 (1985).
\bibitem{soc}
K. Sneppen and M. H. Jensen, Phys.\ Rev.\ Lett.\ {\bf 70}, 3833 (1993); {\bf 71}, 101 (1993); Phys.\ Rev.\ E {\bf 49}, 919 (1994); P. Bak and K. Sneppen, Phys.\ Rev.\ Lett.\ {\bf 71}, 4083 (1993); M. Paczuski, S. Maslov, and P. Bak,  Europhys.\ Lett.\ {\bf 27}, 97 (1994). 
\bibitem{red_81&82a}
S. Redner, J.\ Phys.\ A:\ Math.\ Gen.\ {\bf 14}, L349 (1981);  Phys.\ Rev.\ B {\bf 25}, 3242 (1982).
\bibitem{red_83}
S. Redner, in {\em Percolation Structures and Processes}, edited by G. Deutscher {\em et al}. (Adam Hilger, Bristol, 1983).
\bibitem{perc}
RDNs were already contained implicitly in the pioneering work of S. R. Broadbent and J. M. Hammersley on percolation, Proc.\ Philos.\ Soc.\ {\bf 53}, 629 (1957).
\bibitem{janssen_stenull_2000}
H. K. Janssen and O. Stenull, Phys.\ Rev.\ E {\bf 62}, 3173 (2000).
\bibitem{janssen_stenull_directedLetter_2000}
H. K. Janssen and O. Stenull, Phys.\ Rev.\ E {\bf 63}, 025103(R) (2001).
\bibitem{stenull_janssen_directedResistance_2000}
O. Stenull and H. K. Janssen, to appear in J.\ Stat.\ Phys.
\bibitem{kenkel_straley_82} 
See also S. W. Kenkel and J. P. Straley, Phys.\ Rev.\ Lett.\ {\bf 49}, 767 (1982).
\bibitem{blumenfeld_aharony_85} 
R. Blumenfeld and A. Aharony, J.\ Phys.\ A {\bf 18}, L443 (1985).
\bibitem{stephen_78}
M. J. Stephen, Phys.\ Rev.\ B {\bf 17}, 4444 (1978).
\bibitem{harris_87}
A. B. Harris, Phys.\ Rev.\ B {\bf 35}, 5056 (1987).
\bibitem{harris_lubensky_87}
See, e.g., A. B. Harris and T. C. Lubensky, Phys.\ Rev.\ B {\bf 35}, 6964 (1987).
\bibitem{cardy_sugar_80}
J. L. Cardy and R. L. Sugar, J.\ Phys.\ A {\bf 13}, L423 (1980).
\bibitem{janssen_81}
H. K. Janssen, Z.\ Phys.\ B {\bf 42}, 151 (1981).
\bibitem{janssen_2000}
H. K. Janssen, J. Stat. Phys. {\bf 103}, 801 (2001).
\bibitem{janssen_dynamic}
H. K. Janssen, Z.\ Phys.\ B {\bf 23}, 377 (1976); R. Bausch, H. K. Janssen, and H. Wagner, {\em ibid}. {\bf 24}, 113 (1976); H. K. Janssen, in {\em Dynamical Critical Phenomena and Related Topics}, edited by C. P. Enz, Lecture Notes in Physics, Vol.~104 (Springer, Heidelberg, 1979).
\bibitem{deDominicis&co}
C. De Dominicis, J.\ Phys.\ (Paris) Colloq.\ {\bf 37}, C-247 (1976); C. De Dominicis and L. Peliti, Phys.\ Rev.\ B {\bf 18}, 353 (1978).
\bibitem{janssen_92} 
H. K. Janssen, in {\em From Phase Transitions to Chaos}, edited by G. Gy\"{o}rgyi, I. Kondor, L. Sasv\'{a}ri, and T. T\'{e}l (World Scientific, Singapore, 1992).
\bibitem{amit_zinn-justin} 
See, e.g., D. J. Amit, {\em Field Theory, the Renormalization Group, and Critical Phenomena} (World Scientific, Singapore, 1984); J. Zinn-Justin, {\em Quantum Field Theory and Critical Phenomena} (Clarendon, Oxford, 1989).
\bibitem{stenull_janssen_oerding_99}
O. Stenull, H. K. Janssen, and K. Oerding, Phys.\ Rev.\ E {\bf 59}, 4919 (1999).
\bibitem{janssen_stenull_oerding_99}
H. K. Janssen, O. Stenull, and K. Oerding, Phys.\ Rev.\ E {\bf 59}, R6239 (1999).
\bibitem{janssen_stenull_99}
H. K. Janssen and O. Stenull, Phys.\ Rev.\ E {\bf 61}, 4821 (2000).
\bibitem{stenull_2000}
O. Stenull, {\em Renormalized Field Theory of Random Resistor Networks}, Ph.D. thesis, Universit\"{a}t D\"{u}sseldorf, (Shaker, Aachen, 2000).
\bibitem{stenull_janssen_2000a}
O. Stenull and H. K. Janssen, Europhys.\ Lett.\ {\bf 51}, 539 (2000).
\bibitem{kertez_vicsek_94}
See, e.g., J. Kert\'{e}sz, and T. Vicsek, in {\em Fractals in Science}, edited by A. Bunde and S. Havlin (Springer, Berlin, 1994).
\bibitem{coniglio_81&82} 
A. Coniglio, Phys.\ Rev.\ Lett.\ {\bf 46}, 250 (1981); 
J.\ Phys.\ A {\bf 15}, 3829 (1982).
\bibitem{hede&co_91}
See, e.g., B. Hede, J. Kert\'{e}sz, and T. Vicsek, J.\ Stat.\ Phys.\ {\bf 64}, 829 (1991).
\bibitem{arora&co_83}
B. M. Arora, M. Barma, D. Dhar, and M. K. Phani, J.\ Phys.\ C {\bf 16}, 2913 (1983).
\bibitem{sneppen&co_95}
G. Huber, M. H. Jensen, and K. Sneppen, Phys.\ Rev.\ E {\bf 52}, R2133 (1995).
\end{references}
\end{document}